\documentclass[review]{elsarticle}

\usepackage{lineno,hyperref}
\usepackage{adjustbox}
\modulolinenumbers[5]
\usepackage{dsfont}
\usepackage{amsmath}
\usepackage[utf8]{inputenc}
\usepackage{xcolor}
\newcommand{\sys}[1]{\textsc{#1}}

\journal{Journal of \LaTeX\ Templates}

\begin{document}

\begin{frontmatter}

\title{A Machine Learning Model for Nowcasting Epidemic Incidence}


\author[mymainaddress]{Saumya Yashmohini Sahai
\hspace{1.0mm}\href{https://orcid.org/0000-0002-4254-5223}{\includegraphics[width=3mm]{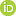}}}
\ead{sahai.17@osu.edu}

\author[mymainaddress]{Saket Gurukar
\hspace{1.0mm}\href{https://orcid.org/0000-0002-1699-5714}{\includegraphics[width=3mm]{ORCID-iD_icon-16x16}}
}
\author[mysecondaryaddress]{Wasiur R. KhudaBukhsh\hspace{1.0mm}\href{https://orcid.org/0000-0003-1803-0470}{\includegraphics[width=3mm]{ORCID-iD_icon-16x16}}}
\author[mymainaddress]{Srinivasan Parthasarathy
\hspace{1.0mm}\href{https://orcid.org/0000-0002-6062-6449}{\includegraphics[width=3mm]{ORCID-iD_icon-16x16}}}

\author[mysecondaryaddress]{Grzegorz A. Rempa{\l}a\hspace{1.0mm}\href{https://orcid.org/0000-0002-6307-4555}{\includegraphics[width=3mm]{ORCID-iD_icon-16x16}}\corref{mycorrespondingauthor}}
\cortext[mycorrespondingauthor]{Corresponding author}
\ead{rempala.3@osu.edu}

\address[mymainaddress]{Department of Computer Science and Engineering \\ The Ohio State University}
\address[mysecondaryaddress]{Division of Biostatistics and Mathematical Biosciences Institute \\ The Ohio State University}

\begin{abstract} Due to delay in reporting, the daily national and statewide COVID-19 incidence counts are often unreliable and need to be estimated from recent data. This process is known in economics as nowcasting. We describe in this paper  a simple random forest statistical model for nowcasting the  COVID - 19 daily new infection counts based on historic data along with a set of simple covariates, such as the currently reported infection counts, day of the week,  and time since first reporting. We apply the model to adjust the daily infection counts in Ohio, and show that the predictions from this simple data-driven method compare favorably both in quality and computational burden to those obtained from the state-of-the-art hierarchical Bayesian model employing a complex statistical algorithm.
\end{abstract}


\end{frontmatter}


\section{Introduction}

\begin{figure}
    \centering
    \includegraphics[width=0.75\linewidth]{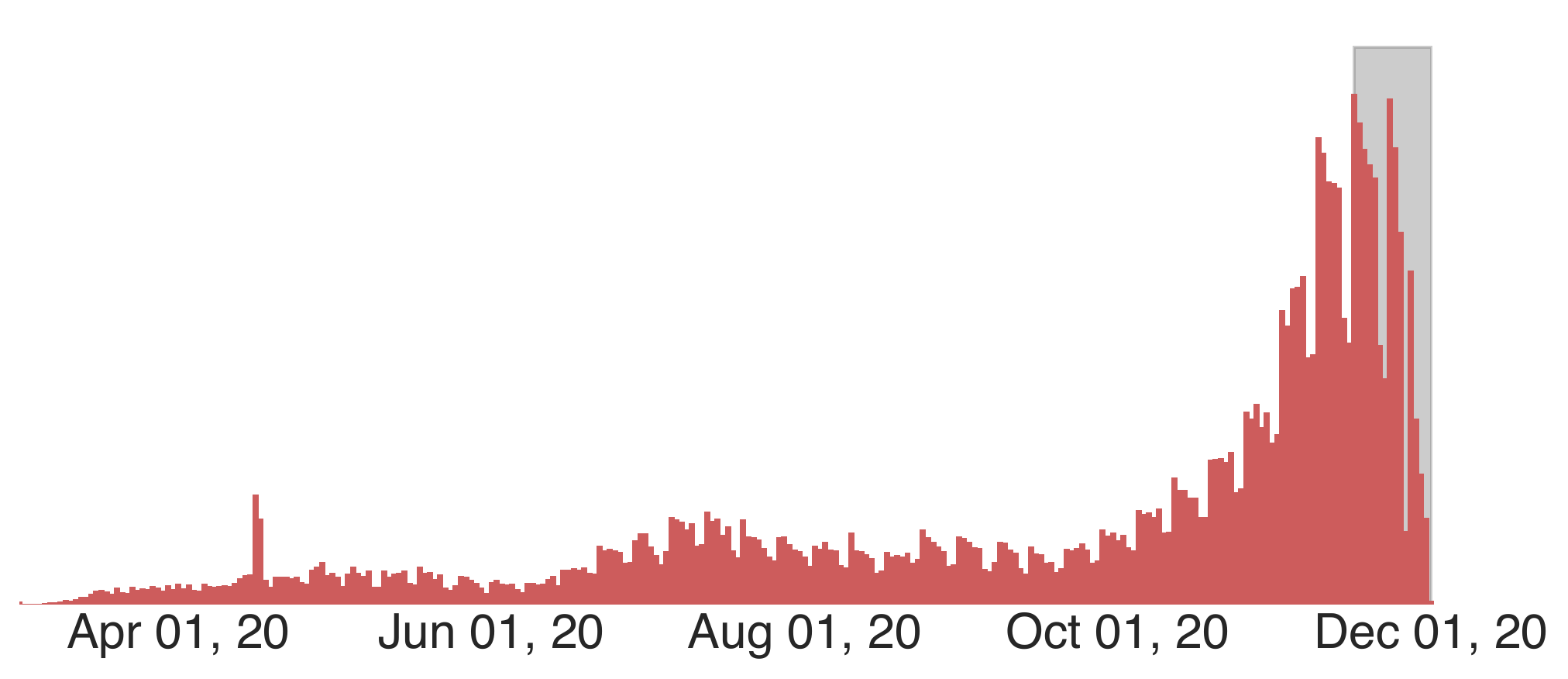}
    \caption{Daywise COVID-19 cases in Ohio, as on 12-01-2020. 
    The shaded area -- comprising of 21 days -- is the preliminary case data and is likely under-reported to the ODH due to delayed reporting.}
    \label{fig:dailycases}
\end{figure}

The SARS-CoV-2 virus, first observed in the United States (USA) in January 2020 \cite{bedford2020cryptic, CDCfirst, fauver2020coast}, is highly contagious \cite{WHOTransmission} and has spread in both urban and rural regions \cite{paul2020progression, mueller2021impacts} of the USA.  To gauge and combat the SARS-CoV-2 spread, governments and health 
organizations have set up public information systems such as COVID-19 dashboards \cite{ODHCovid, NYCCovid, CaliCovid, UtahCovid}. 
These dashboards are useful to brief the public \cite{NYCCovid}  about the current state of COVID-19 in specific regions, make 
data-driven public health decisions \cite{UtahCovid}, and improve transparency in governance  \cite{fell2020trust}. Many of these dashboards show the number of daily new  infections (daily incidence), where the infection count on a particular date refers to the number of people who started experiencing disease symptoms on that date (i.e., the onset date of illness). Whereas reporting onset dates is very useful from the viewpoint of contact tracing and disease spread monitoring, it is also challenging due to unavoidable delays. \cite{harris2020overcoming, greene2021nowcasting, ODHCovid}. These delays are  often due to the time-lags between experiencing initial symptoms and  seeking care, receiving testing results, and  updating the statewide records  \cite{wys, odhantigen}. As a consequence,   the  incidence reporting  based on onset counts leads to under-counting of the present and most recent cases. Dashboards often explicitly warn about this  problem   \cite{WHOdash, WashingtonCovid}. Figure~\ref{fig:dailycases} shows one such example from 
COVID-19 Dashboard maintained by the Ohio Department of Health (ODH) \cite{ODHCovid} where the region of possible under-reporting is marked with a grey rectangle.

The incomplete  current count data poses huge challenges for both local and national healthcare  policymakers as they strive to make difficult public health decisions (e.g., introduce lockdowns, curfews, evaluate vaccination effects, etc) in real time to limit the spread of the virus. The use  of  statistical methods to moderate the effects of incomplete data could   help 
reduce uncertainty in public health decision-making during the COVID-19 pandemic and increase public awareness of the most recent disease trends.  

While forecasting COVID-19 cases is typically concerned with predicting the future burden of the epidemic, {\em nowcasting} \cite{van2019nowcasting, mcgough2020nowcasting, lawless1994adjustments} addresses the problem of delayed reporting and focuses on the estimation of current case counts from not-too-distant historic data. Given the under-reported infection data for a particular date, the nowcasting models estimate the total number of current infections for that date, which will be reported eventually. In the literature, there exist several sophisticated statistical methods for addressing the issue of nowcasting for COVID-19. For instance, Wu et al. \cite{wu2020nowcasting}  nowcast the probable size of the COVID-19 outbreak in Wuhan, China. The authors estimate the basic reproduction number $R_0$ from their proposed non-homogeneous counting process modeling the exported number of international cases from Wuhan and the global human mobility data from/to Wuhan. The authors then used the estimated $R_0$ in the Susceptible-Exposed-Infected-Recovered or   SEIR model \cite{aron1984seasonality} for nowcasting and forecasting the outbreak's size. The nowcasting problem for delayed reporting of COVID-19 cases is also addressed by Silva et al. \cite{silva2020population} and Greene et al. \cite{greene2021nowcasting} using Bayesian smoothing approach \cite{mcgough2020nowcasting} where the authors model the delayed number of reported cases with their proposed Markov counting processes.

 In this paper, we propose a simple yet efficient machine learning model that addresses the problem of nowcasting in a way that is easily understood by non-experts and therefore suitable for presenting to public health decision-makers. The only data our proposed model requires can be readily collected  from publicly available dashboards. Despite its simplicity, the model is seen to predict, with high accuracy (measured with the typical regression-style  $R^2$ value), the number of people who start experiencing COVID-19 symptoms on a particular date. We also show that our proposed model outperforms the state-of-the-art hierarchical Bayesian model \cite{kline2021bayesian} in terms of nowcasting accuracy while being  also approximately 72000x faster. Our model predictions can also be utilized as input to other forecasting models, for instance, the  ones created for ODH \cite{IDIModel} that forecast the future number of infections and subsequent hospital burden in Ohio.



\section{Materials and Methods}

\subsection{Data processing}
\label{subsec:datapreprocessing}
To perform our analysis, we used the public data available at ODH COVID-19 dashboard\footnote{https://coronavirus.ohio.gov/wps/portal/gov/covid-19/dashboards/overview}, which is updated daily. It provides the daily partial incidence count, that is, the count of all   individuals $i_{td}$ reported on a given day $t$ to be  confirmed  COVID-19 cases with the day of onset $d$ where  $d\le t$. For our analysis we  aggregated cases by the onset date to get the state-level progression of the onset reporting. Accordingly, the infection count $I_{TD}$ on  a specific day $T$ for a given specific onset date $D$ is given by

\begin{equation}
    I_{TD} = \sum_{ d\le t\le T}  \mathds{1}_{i_{td}},
    \label{eq:rawinfection}
\end{equation}

where $ \mathds{1}_{i_{td}}$ is the indicator function 

\begin{equation}
   \mathds{1}_{i_{td}}=\begin{cases}
               1 \text{ where }t\le T, d=D,\\
              0 \text{ otherwise.}
            \end{cases}
\end{equation}

\begin{figure}[t]
\includegraphics[width=\textwidth]{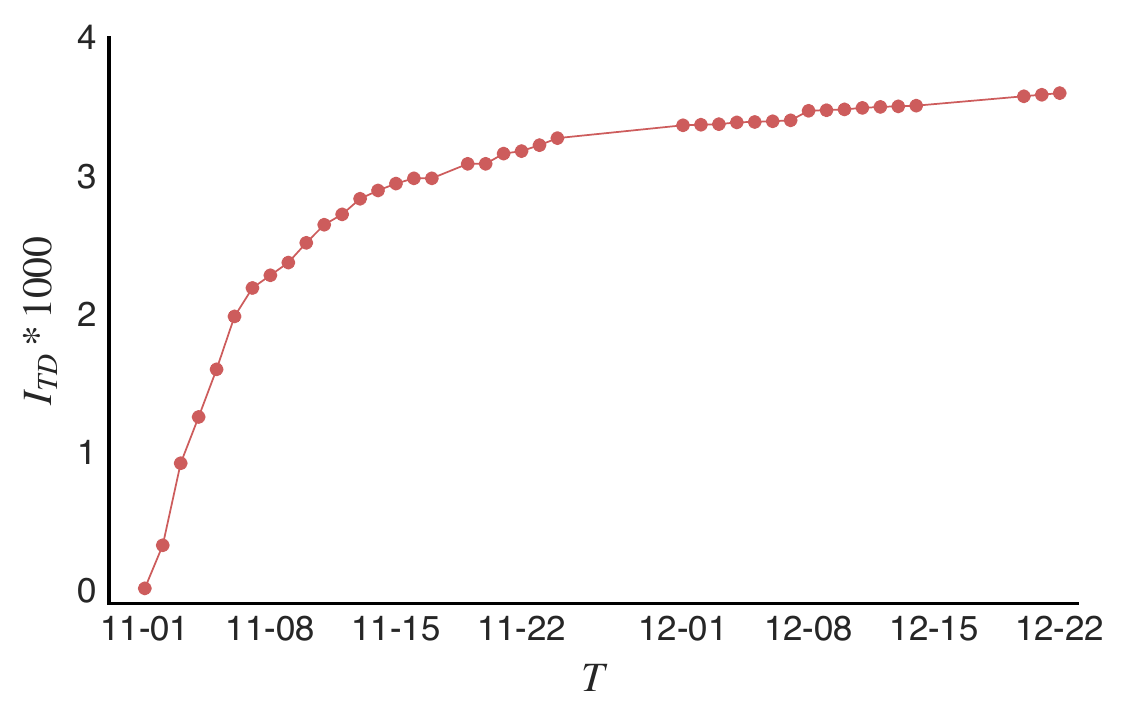}
\caption{Progression of infection count $I_{td}$ for a specific $d$ value (11-01-2020) over $t$ ranging from 11-01-2020 to 12-22-2020. There is a steep rise in the infection count in the initial days of data collection as the data is backfilled, but it gradually stabilizes.} \label{fig:rawtrend}
\end{figure}

Note that for  a given $D$,  $I_{TD}$ is non-decreasing as a function of $t$ and, assuming that it is also bounded,  it has a limit as $ T\to \infty$. This is illustrated in Figure~\ref{fig:rawtrend} where we see that over the course of 52 days   $I_{TD}$ becomes  approximately  a constant.

We denote the asymptotic stable value of $I_{TD}$ for an onset date $D$ by
\begin{equation}
    I^{s}_{D} = \lim_{T\to \infty} I_{TD}
\end{equation}
and  define  $F_{TD}$ as the amount of undercounting for a specific $D$ on day $T$  given by 
\begin{equation}
\label{eq:fractionequation}
    F_{TD} = 1 - \frac{I_{TD}}{I^{s}_{D}}.
\end{equation}
We may think about $F_{TD}$ as a standardized  measure  of undercounting that is also   robust  to changes in incidence rates during the course of the pandemic. In what follows, we therefore consider   $F_{TD}$ in place of $I_{TD}$.  The plot of  $F_{TD}$  as a function of time $T$  is presented in Figure \ref{fig:rawtrend}.  Note that although  in general    $F_{TD}\to 0$ as $T\to \infty$,   we may see clearly  from the plot  that   this convergence is not necessarily monotone and that in the fixed  time window   $I_{TD}$ only approximately stabilizes as it approaches $I^{s}_{D}$.  In order to improve data stability in the time windows of interest,  we consider the $I_{TD}$ limit to be reached in practice as soon as  $F_{TD} < 0.05$. This  particular cutoff value was chosen by cross-validation  $[0, 0.5]$, as described in Section \ref{subsec:model}.  

In order to cross-validate and measure the prediction testing error,  data to be used for nowcasting is split into a training and a validation (testing) set based on $t$, where all $F_{td}$ with $t < T_{train}$ are in the former and $t > T_{train}$ are in the latter.

\subsection{Model}
\label{subsec:model}
\paragraph{Covariates}

The model includes the following features to predict the $F_{td}$.
\begin{itemize}
    \item \underline{Days since data collection} ($\Delta$). For any given infection count ${I_{td}}$ reported on day $t$ with onset date $d$, we define this feature as
    \begin{equation}
        \Delta_{td} = t - d.
    \end{equation}
    \item  \underline{Day of the week} ($\omega_t$).  This categorical variable denotes the day of the week for $t$, at which data is being reported, $\omega_t \in \{$Mo, Tu, We, Th, Fr, Sa, Su$\}$.
    \item \underline{Raw infection count} ($I_{td}$).  This is the daily partial incidence count for the pandemic, as described in equation \ref{eq:rawinfection}.
\end{itemize}

\paragraph{Random forest regression} We train a random forest (RF) regression model \cite{randomforests} on the data partition defined in section \ref{subsec:datapreprocessing}, to predict $F_{td}$ from the covariates. Formally, we may write 
\begin{equation}
    F_{td} = f(\Delta_{td}, \Delta_{td}^2, \Delta_{td}^3, \omega_i, I_{td}),
    \end{equation}

where $f$ is the RF  model.


\newcommand{\mf}{$F_{td}$ }
\newcommand{\ds}{$\Delta_{td}$ }

\section{Results}

\begin{figure}
\includegraphics[width=\textwidth]{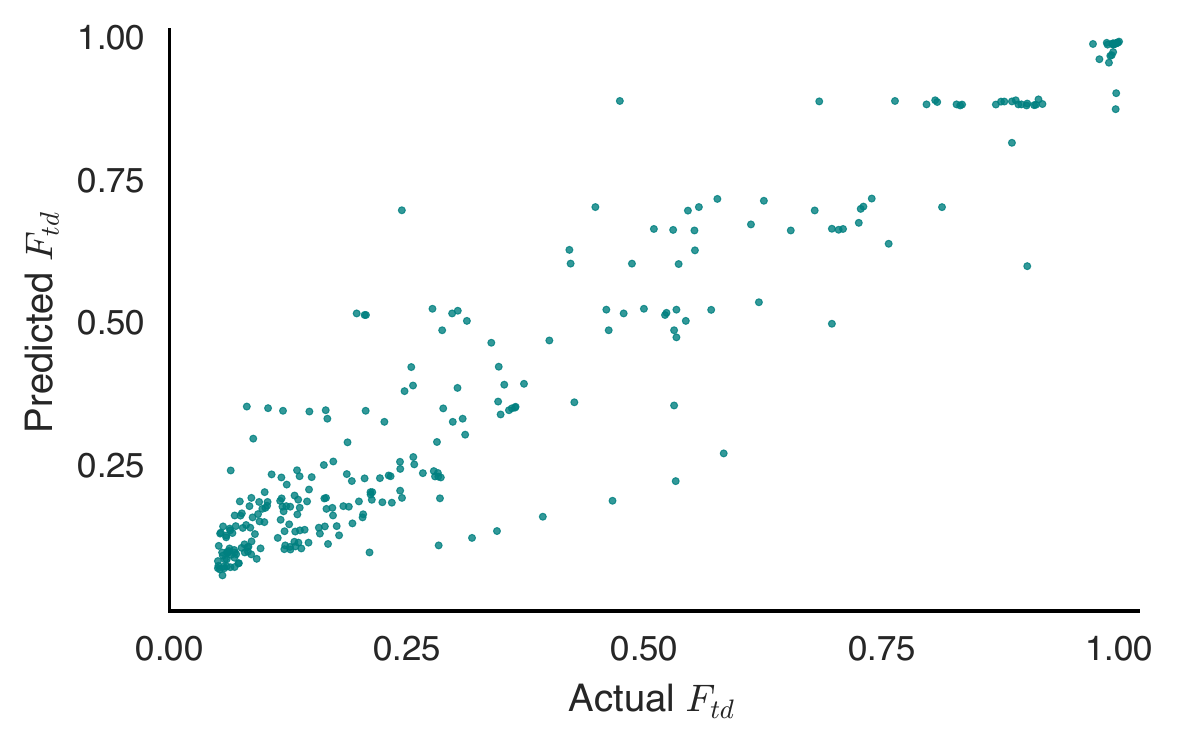}
\caption{Actual vs Predicted $F_{td}$ on the testing dataset. Robust prediction of \mf is crucial for correct prediction of final infection count $I^s_d$.} \label{fig:predvsact}
\end{figure}

\paragraph{Goodness of fit} The explained variance ($R^2$ value) is used to evaluate the goodness of fit of the model on both the training data  (time window from 10-01-2020 to 11-15-2020) and on the  testing data (time window from 11-16-2020 to 12-15-2020). The predictions from the fitted model plotted against the true values in test data can be seen in Figure~\ref{fig:predvsact}. The explained variance is $0.99$ on the training  data and $0.89$ on the testing data, which shows that the model's prediction of $F_{td}$ generalizes well to the unseen data. 

\paragraph{Importance of covariates} The relative importance of covariates (the Gini importance or the mean decrease in impurity) in the fitted model, described in Section~\ref{subsec:model} can be seen in Table \ref{tab:feature_importance}. The covariate \emph{days since data collection} ($\Delta$), along with its quadratic and cubic transforms turn out to be the most important features in determining the fraction of missing data \mf. The \emph{day of the week} $\omega_i$ has much less relative importance.


\begin{table}[]
    \centering
    \begin{adjustbox}{width=\textwidth}
    \begin{tabular}{c|c|c|c|c|c|c|c|c|c|c|c}
        Covariate & $\Delta^2$ & $\Delta^3$ & $\Delta$ & $I$ & $\omega_t=\text{Th}$ & $\omega_t=\text{Tu}$ & $\omega_t=\text{We}$ & $\omega_t=\text{Fr}$ & $\omega_t=\text{Mo}$ & $\omega_t=\text{Sa}$ & $\omega_t=\text{Su}$ \\
        \hline
        Importance &  0.337 & 0.325 & 0.311 & 0.013 & 0.003 & 0.003 & 0.003 & 0.002 & 0.001 & 0.001 & 0.001 \\
    \end{tabular}
    \end{adjustbox}
    \caption{Relative (Gini) importance of covariates. Days since data collection ($\Delta$) and its transformations are the most important, with day of the week $(\omega_t)$ having the least effect.}
    \label{tab:feature_importance}

\end{table}

\begin{figure}
\includegraphics[width=\textwidth]{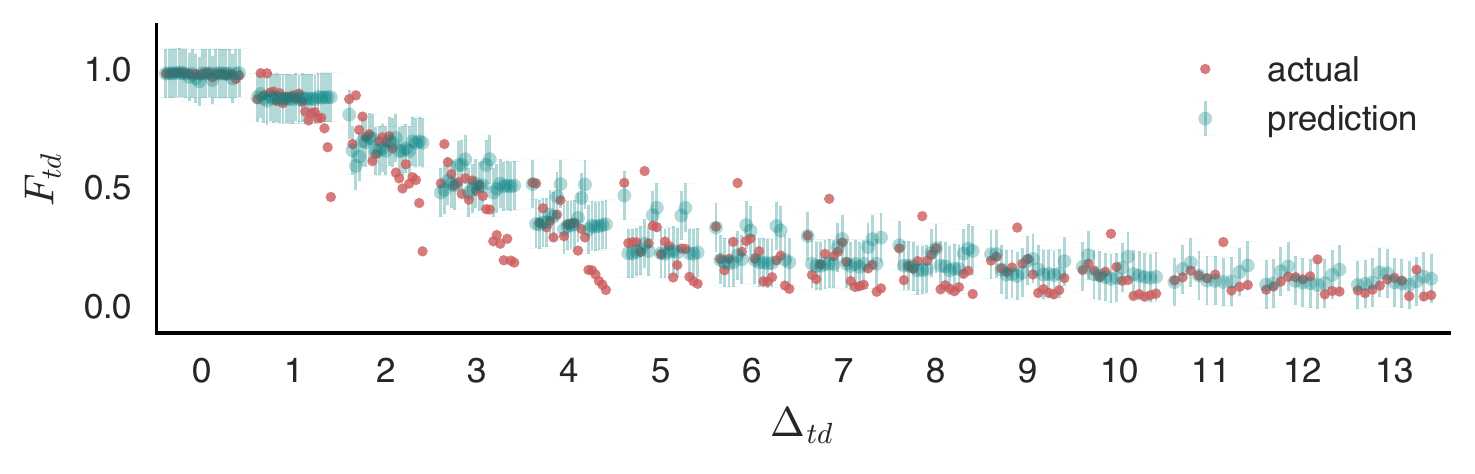}
\caption{Predicted missing fraction, \mf  at various \ds } \label{fig:predicted_missingness}
\end{figure}

\paragraph{Prediction of missingness $F_{td}$} Figure \ref{fig:predicted_missingness} shows the prediction of $F_{td}$ for  different values of $\Delta_{td}$. As seen from the plot, the model predictions are close to the true $F_{td}$ when $\Delta > 4$. The good agreement  at $\Delta = 0$ is trivial, as at first date of collection, $F_{td}$ is almost always close to $1.0$ and thus easy to predict. It is also evident that first 3-4 days of data collection seem to be unreliable in predicting the correct \mf and therefore should be utilized cautiously  in the nowcasting predictions.

\begin{figure}
    \centering
  \begin{minipage}{0.5\textwidth}
    \includegraphics[width=6.3cm]{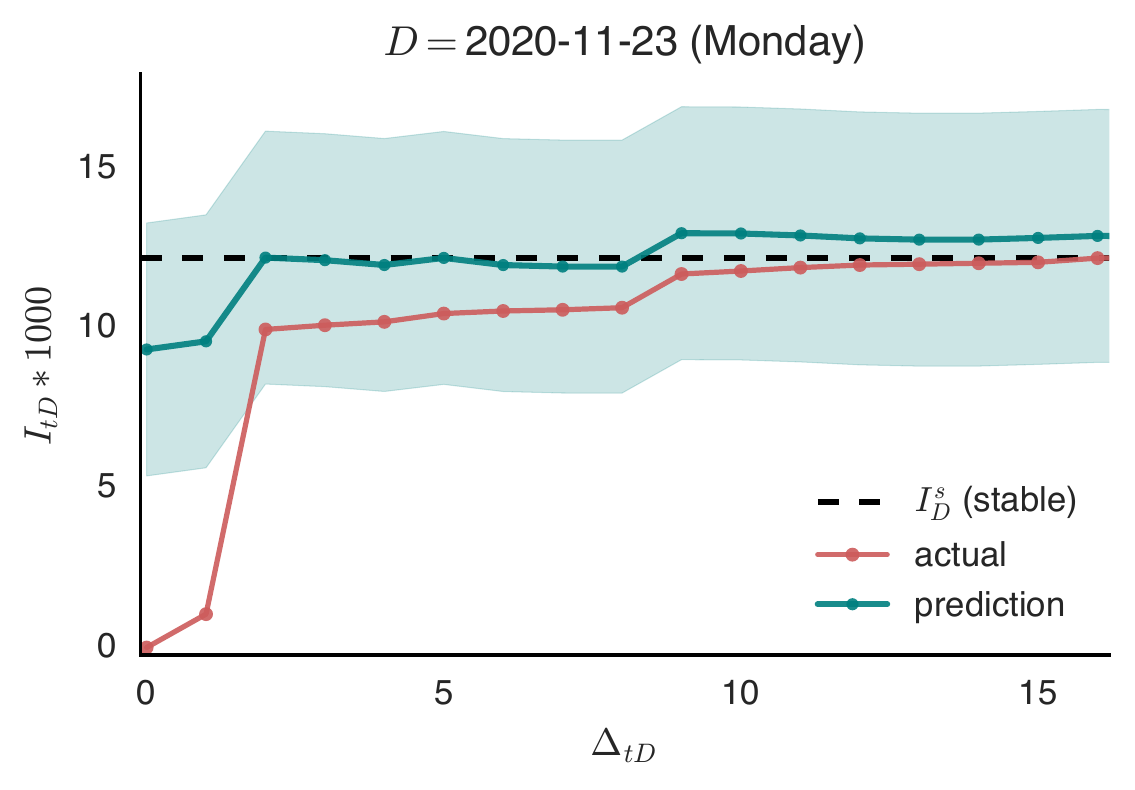}
    
  \end{minipage}
  \hfill
  \centering
  \begin{minipage}{0.49\textwidth}
    \includegraphics[width=6.3cm]{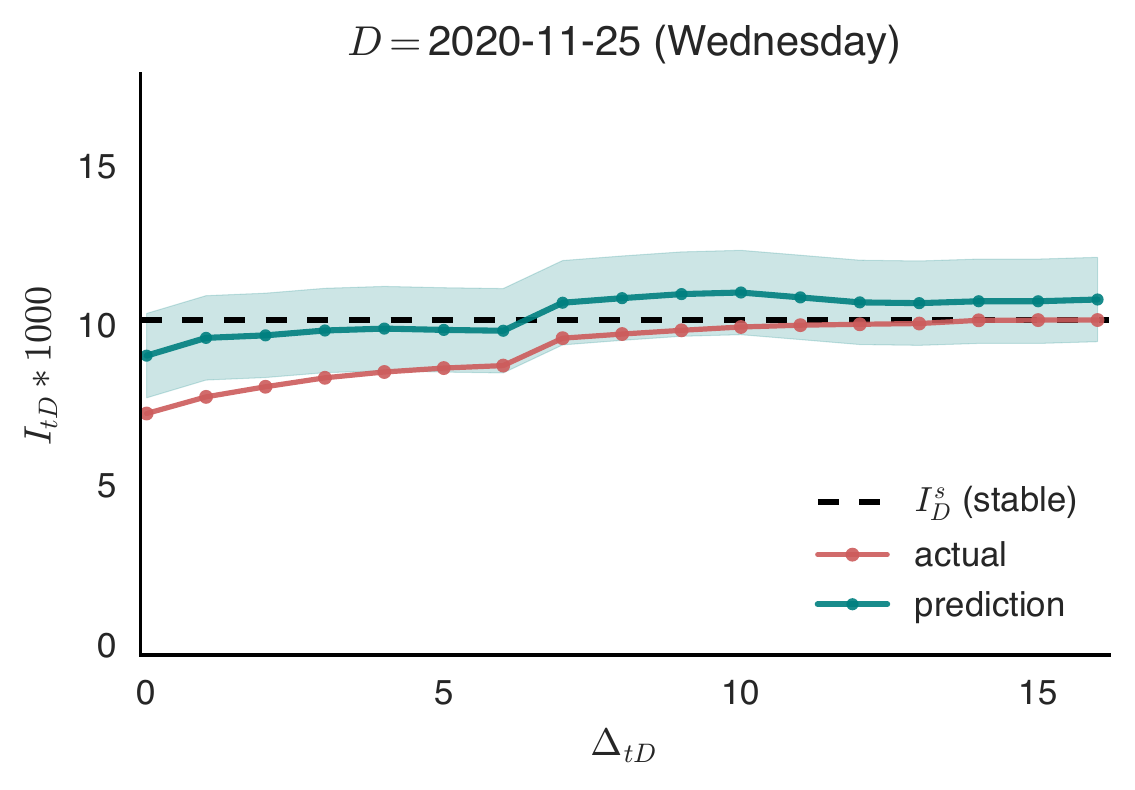}
   
  \end{minipage}
  
  \centering
  \begin{minipage}{0.5\textwidth}
    \includegraphics[width=6.3cm]{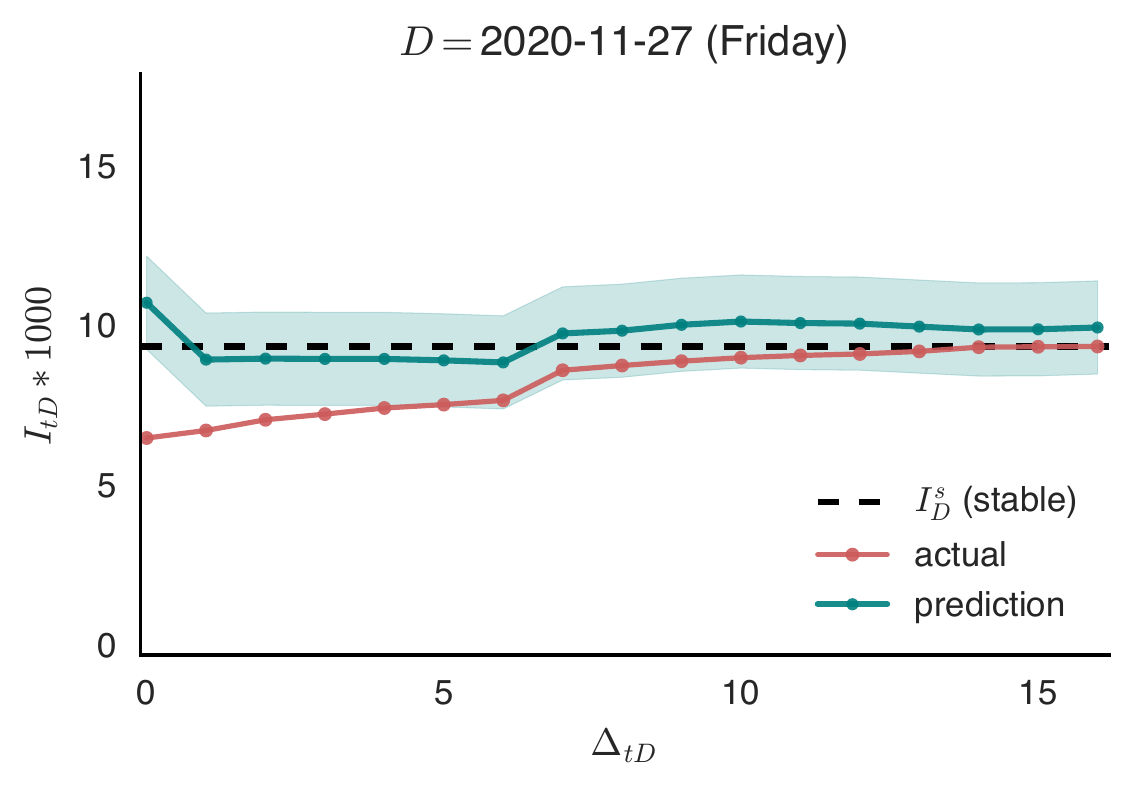}
    
  \end{minipage}
  \hfill
  \centering
  \begin{minipage}{0.49\textwidth}
    \includegraphics[width=6.3cm]{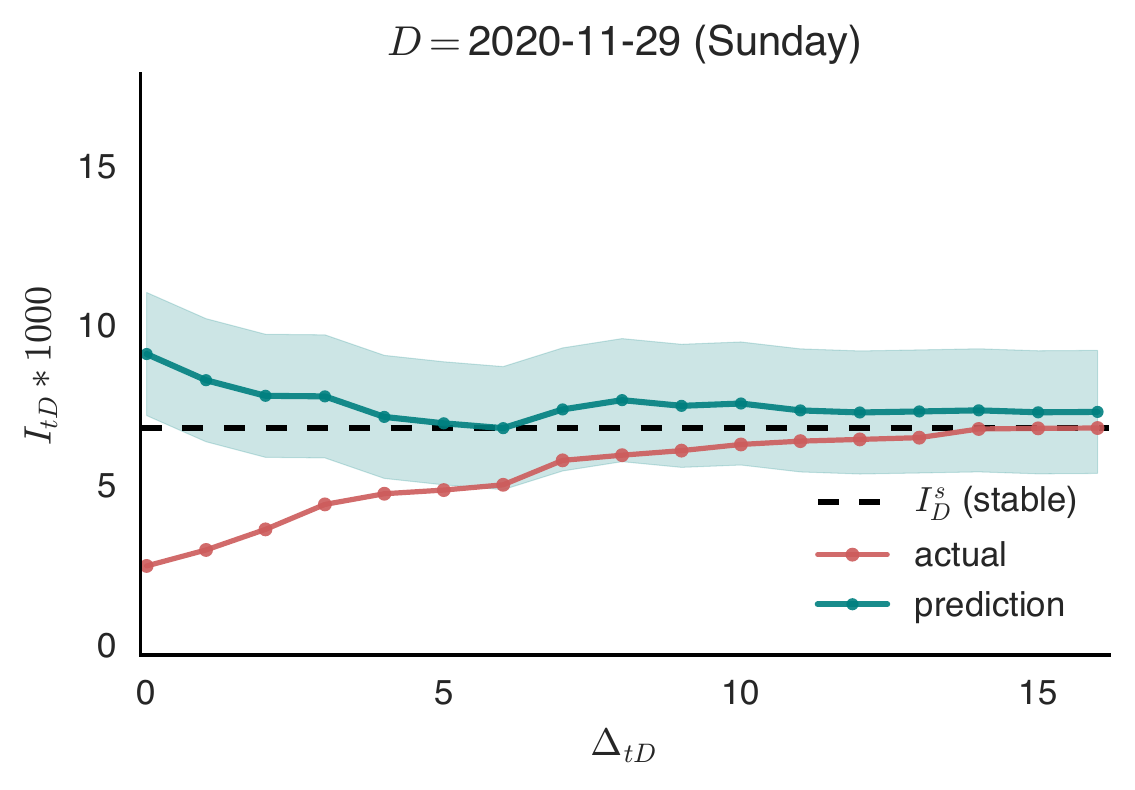}
   
  \end{minipage}
    \caption{Prediction of raw infection count}
    \label{fig:infectionpred}
    \end{figure}

\paragraph{Actual count prediction} Based on the prediction of $F_{td}$ and the current observed count $I_{td}$, we use  \eqref{eq:fractionequation}, to get the estimate of $I^s_d$, which is the stable value of the infection count on day $d$. The typical trends for 4 different days of the week can be seen in Figure~\ref{fig:infectionpred}. The infection count from the model predicts the stable value $I^s_d$ robustly after five days (starting from $\Delta = 5$), and in some cases even earlier. In Figure~\ref{fig:infectionpred} we may see that irrespective of the day of the week (Monday, Wednesday, Friday, Sunday), the model is seen  to predict  the  value of  $I^s_d$ with good accuracy. We may also note  that on Monday and Sunday the model predictions have  higher uncertainty likely  due to the effect of weekend test processing slowdown.

\begin{figure}
\includegraphics[width=\textwidth]{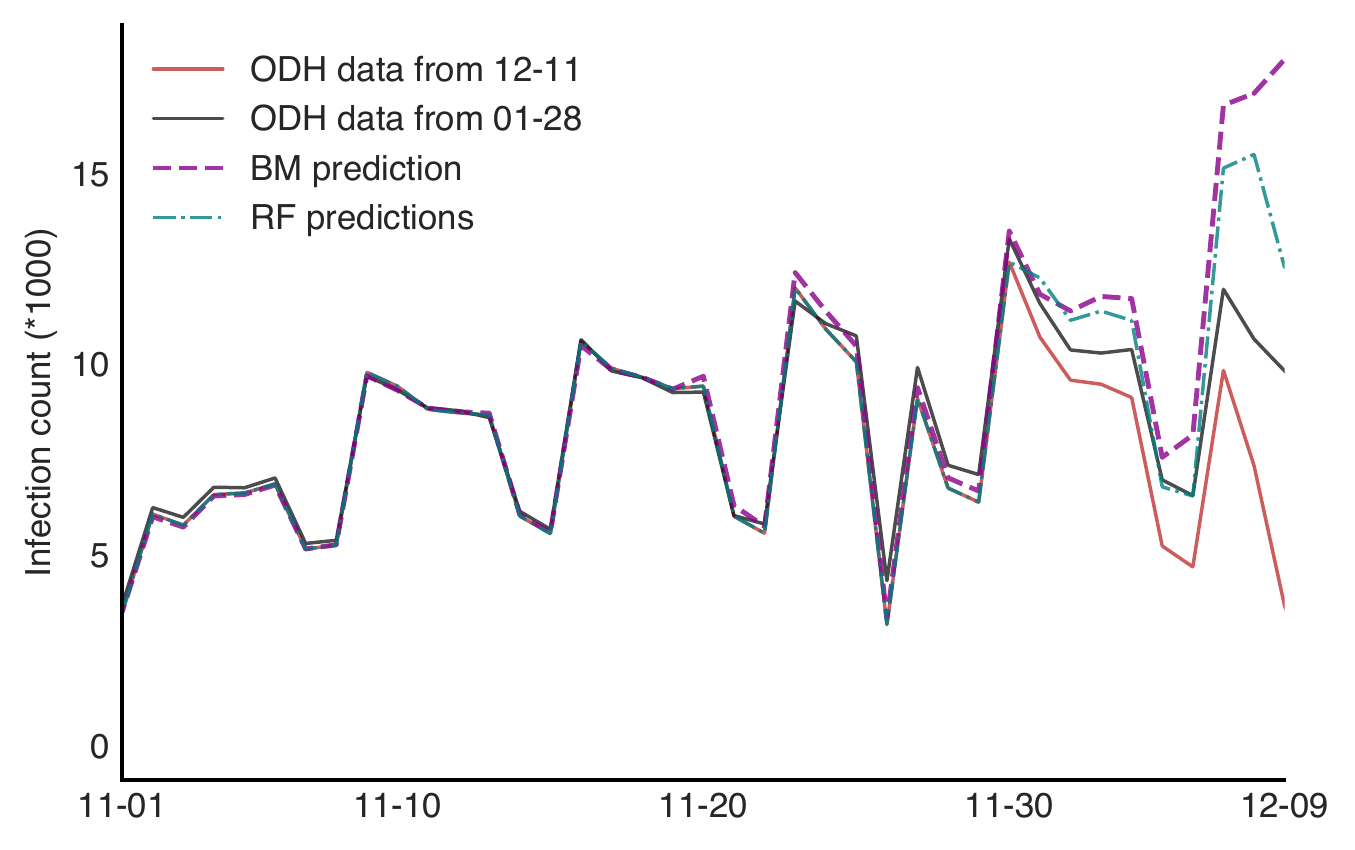}
\caption{Comparison between the Bayesian model (BM) and random forest (RF) model from 11-01-2020 to 12-09-2020. } \label{fig:}
\end{figure}

\begin{figure}
\includegraphics[width=\textwidth]{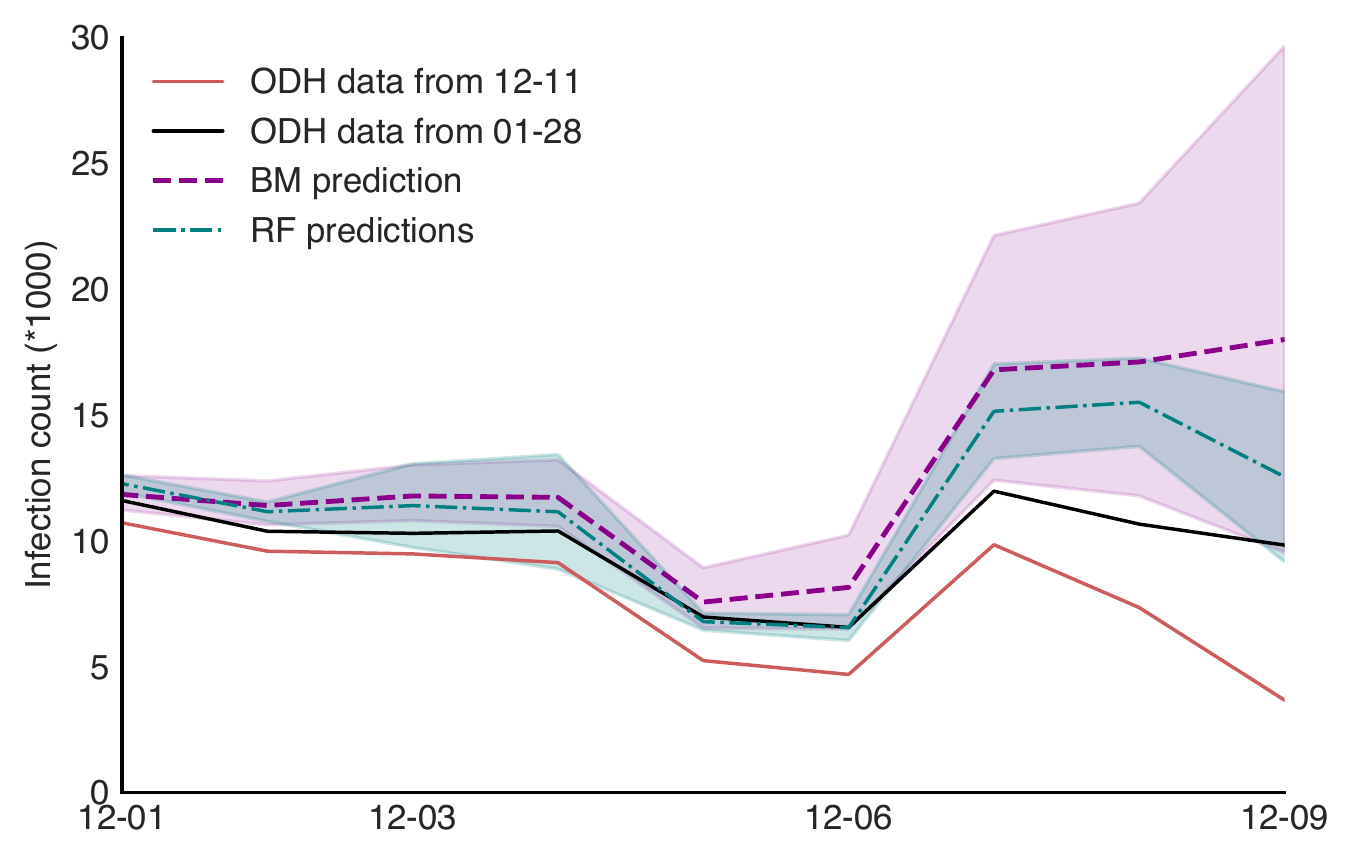}
\caption{Comparison between the Bayesian model (BM) and random forest (RF) model from 12-01-2020 to 12-09-2020. } \label{fig:}
\end{figure}

\paragraph{Comparison with the Bayesian model}
In order to provide some context for assessing the quality of the RF model predictions,  we compare our results with a state-of-the-art hierarchical Bayesian model proposed recently by Kline et al. \cite{kline2021bayesian}, which has been used for the same purpose of nowcasting COVID-19 cases in the state of Ohio. The model, which we   refer to as the Bayesian model (BM) in the following, is more elaborate than ours as  it has also  a spatial component. Specifically, it  keeps track of COVID-19 cases over time in different geographical regions (counties in Ohio). Although in our comparison we aggregate BM spatial counts, for the sake of completeness we briefly describe here the entire  model  along with  its spatial component.   Denoting  by  $Y_{i, t}$  the true count of cases in county~$i$ with onset date~$t$  the BM assumes the following Poisson model for the dynamics of the disease:
\begin{align}
    Y_{i,t} &{} \sim \sys{Poisson}\left( \exp\left(O_i + \alpha_{i,t} + X_t \eta_i \right) \right), 
\end{align}
where $O_i$ is an offset of the logarithm of population of county~$i$, the spatio-temporal random variables $\alpha_{i,t}$ are the latent states of the process, the design vector $X_t$ indicates the day of the week, and the vector $\eta_i$ captures the day of the week effect. It is  assumed that  $Y_{i, t}$ is only partially observed for time $t > T_{\text{max}} -D$, where $T_{\text{max}}$ stands for the last onset date and $D$ (assumed $30$ in \cite{kline2021bayesian}) is the maximum reporting delay following onset. BM also  uses a semi-local linear trend model \cite{Brodersen2015Causal} for the spatio-temporal random variables $\alpha_{i,t}$. Further, the spatial correlation is accounted for using an intrinsic conditional auto-regressive model. The reporting delay is described by a Multinomial-Dirichlet model as follows. Denoting by  $Z_{i,t,d}$ the  count of cases in county $i$ with onset date $t$, which are observed $d$ days after $t$, one  defines $Z_{i,t} = \left(Z_{i,t,0}, Z_{i,t,1}, \ldots, Z_{i,t,D}  \right)$. 
Then, the Multinomial-Dirichlet model prescribes 
\begin{align*}
    Z_{i,t} & \sim \sys{Multinomial}\left(p_{i, t}, Y_{i, t} \right), \\
    p_{i,t} & \sim \sys{GeneralizedDirichlet}\left(a_{i,t}, b_{i,t} \right), 
\end{align*}
where the vectors $a_{i, t}$ and $b_{i, t}$
are described in terms of mean and dispersion parameters \cite{Stoner2020Hierarchical}. The choice of a Generalized Dirichlet distribution allows for modeling potential overdispersion in $p_{i,t}$ (see \cite{Stoner2020Hierarchical}). Moreover, it leads to a convenient $\sys{Beta-Binomial}$ conditional distribution representation for the components $Z_{i,t,d}$. For the purpose of Bayesian analysis, the authors specify normally distributed priors for the parameters and use the R package \texttt{nimble} to perform a Markov chain Monte Carlo (MCMC) algorithm. The authors report a run time of approximately 20 hours for $30,000$ iterations. 

To compare the two models, we calculate  the $L_2$ distance  between the  predictions made by the RF  and the Bayesian model, respectively  and  the actual known stable values in the Ohio COVID-19 daily counts dataset. We report the ratio of  the two $L_2$ distance  values as a measure of relative closeness of the models to the true (stable) data value for days $T-10$ to $T$ and $T-10$ to $T-5$, where $T$ is the last available date in the data. The results are presented in Table \ref{tab:comparison}. As can be seen in the table, the predictions by the random forest model are relatively closer to the true values than those generated using the Bayesian model estimates. The ratio  is smaller  in the full 10 day window, indicating that the RF model makes better  predictions than BM for days that are  close to data collection.

\begin{table}[]
    \centering
    \begin{tabular}{c|c| c}
         & $T-10:T$ & $T-10:T-5$ \\
         \hline
    $RF/BM$ & 0.565 & 0.726 \\

    \end{tabular}
    \caption{The ratio of $L_2$ norm of nowcasted predictions from the Bayesian model (BM) and random forest model (RF) from the true stable values at two different time instants $T$ and $T-5$. The ratio values below one indicate that in both cases  the RF model performs better than   BM.}
    \label{tab:comparison}
\end{table}

\section{Summary and Discussion}
We  presented here a simple method for nowcasting COVID-19 cases from historic data on  daily incidence of new cases, as measured by the onset of symptoms. Such type of data is now widely available for all states in the USA as well as for most countries in the world. When the need to take immediate decisions on governance or policy arises, nowcasting can be a useful tool in providing more accurate estimates about disease incidence
and spread.  Specifically, our proposed nowcasting algorithm uses a random forest (RF) regression methodology and leverages covariates that are based on day of the week, the number of days passed since first data collection and total incidence so far. 

The proposed algorithm is both  conceptually simple and computationally efficient. Our results also suggest that  it compares favorably with a much more elaborate Bayesian model. We have illustrated the application of our approach on publicly available  data from COVID-19 daily onsets in Ohio, as available from the state's COVID-19 interactive dashboard. We observed that the model is able to predict the final incidence for a day, within 3 to 4 days of data collection. We also find that the number of days passed since first data collection, along with its transformations (or derivatives), are the most important covariates in predicting the final incidence. 

In order to make our  RF method predictions broadly available to interested researchers and practitioners, we have created a publicly available and accessible interactive notebook (see below). As described in the repository, the notebook allows one to use our algorithm to nowcast current COVID-19 onset occurrences, based on any user-provided historic data supplied in appropriate format.  

The problem of nowcasting historic data is an important one, specially during the current COVID-19 pandemic, when delays in reporting can snowball into sub-optimal policies and actions, that can cost lives and create unnecessary societal burden. Our proposed method allows both general public and health providers to carefully monitor the pandemic trends and make informed decisions. The ideas we presented while focused on COVID-19 can be broadly applicable to similar public health problems  in the future.

\section*{Software Availability}
\noindent The interactive notebook for performing the  nowcasting using the random forest approach described in the paper, along with installation instructions, is freely available at \href{https://github.com/sahaisaumya/nowcasting}{https://github.com/sahaisaumya/nowcasting}

\section*{Acknowledgements}
\noindent This research was partially funded by NSF grants  DMS-1853587 and DMS-2027001  to GAR. The work of WKB was supported by the President's Postdoctoral Scholars Program (PPSP) of the Ohio State University.
We would like to thank Harley Vossler for providing helpful feedback on the interactive notebook.
\bibliography{mybibfile}

\end{document}